\documentclass[twoside,11pt]{article}
\usepackage[accepted]{melba}
\usepackage{amsmath,amsfonts}
\usepackage{color}
\usepackage[dvipsnames]{xcolor}
\colorlet{LightRubineRed}{RubineRed!70!}
\colorlet{Mycolor1}{green!10!orange!90!}
\usepackage{graphicx,wrapfig,lipsum}
\usepackage{amsmath}
\usepackage{graphicx}
\usepackage{csquotes}
\MakeOuterQuote{"}
\usepackage{caption}
\usepackage{color}
\usepackage{enumitem}   
\usepackage{amssymb}
\usepackage[boxed, ruled, vlined, linesnumbered]{algorithm2e}
\usepackage{capt-of}
\renewcommand{\L}[0]{\mathcal{L}}
\newcommand{\A}[0]{\mathcal{A}}
\newcommand{\K}[0]{\mathcal{K}}

\newcommand{\ad}[0]{{\rm ad}}
\newcommand{\Diff}[0]{{\rm Diff}}

\melbaheading{2021:017}{https://www.melba-journal.org/papers/2021:017.html}{2021}{1-20}{12/2020}{02/2022}{Wang and Zhang}{}{}

\ShortHeadings{Deep Learning for Regularization Prediction}{Wang and Zhang}
\firstpageno{1}
\title{Deep Learning for Regularization Prediction in Diffeomorphic Image Registration}

\author{\name Jian Wang \email jw4hv@virginia.edu \\  
	\addr$^{1}$ Computer Science, University of Virginia, VA, USA
	\AND
	\name Miaomiao Zhang\email mz8rr@virginia.edu \\
	\addr$^{1}$Electrical and Computer Engineering, University of Virginia, VA,USA\\
	\addr$^{2}$ Computer Science, University of Virginia, VA, USA
}

\begin{document}

\maketitle

\begin{abstract}
This paper presents a predictive model for estimating regularization parameters of diffeomorphic image registration. We introduce a novel framework that automatically determines the parameters controlling the smoothness of diffeomorphic transformations. Our method significantly reduces the effort of parameter tuning, which is time and labor-consuming. To achieve the goal, we develop a predictive model based on deep convolutional neural networks (CNN) that learns the mapping between pairwise images and the regularization parameter of image registration. In contrast to previous methods that estimate such parameters in high-dimensional image space, our model is built in an efficient bandlimited space with much lower dimensions. We demonstrate the effectiveness of our model on both 2D synthetic data and 3D real brain images. Experimental results show that our model not only predicts appropriate regularization parameters for image registration, but also improving the network training in terms of time and memory efficiency. 
\end{abstract}


\section{Introduction}
Diffeomorphic image registration is a fundamental tool for various medical image analysis tasks, as it provides smooth and invertible smooth mapping (also known as a diffeomorphism) between pairwise images. Examples include atlas-based image segmentation~\citep{ashburner2005unified,gao2016image}, anatomical shape analysis based on geometric changes~\citep{vaillant2004statistics,zhang2016low,hong2017fast}, and motion correction in spatial-temporal image sequences~\citep{de2012temporal,liao2016temporal,xing2019plug}. The nice properties of diffeomorphisms keep topological structures of objects intact in images. Artifacts (i.e., tearing, folding, or crossing) that generate biologically meaningless images can be effectively avoided, especially when large deformation occurs. The problem of diffeomorphic image registration is typically formulated as an optimization over transformation fields, such as a free-form deformation using B-splines~\citep{rueckert2006diffeomorphic}, a LogDemons algorithm based on stationary velocity fields (SVF)~\citep{arsigny2006log}, and a large diffeomorphic deformation metric mapping (LDDMM) method utilizing time-varying velocity fields~\citep{beg2005computing}. 

\begin{figure}[b!]
\begin{center}
 \includegraphics[width=1.0\textwidth] {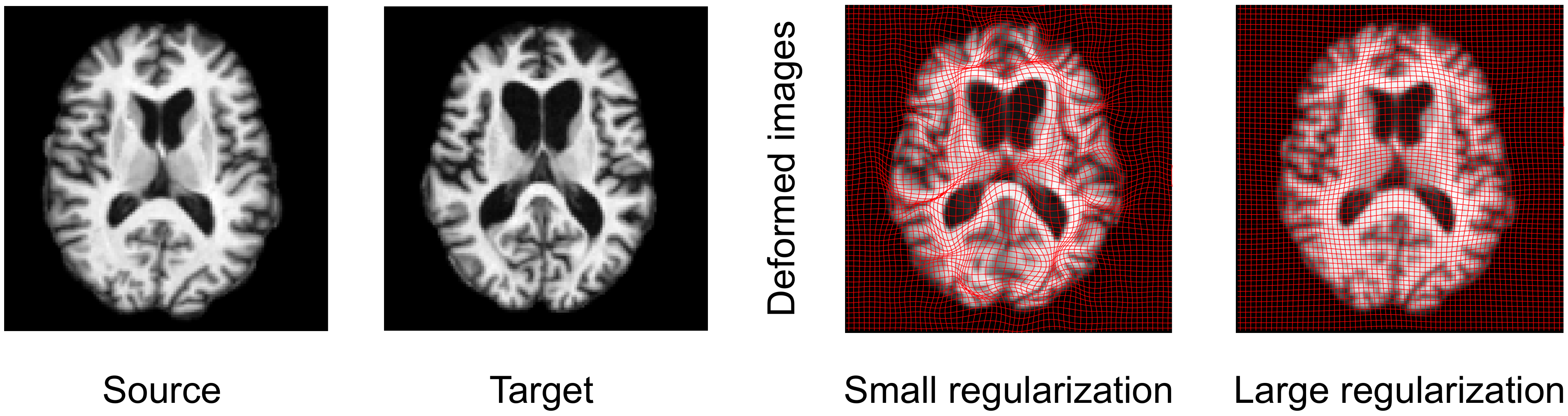}
     \caption{Left to right: examples of transformation fields overlaid with deformed images with under-regularized and over-regularized registration models. A small regularization introduces crossing artifacts on the transformations vs. a large regularization discourages sufficient transformations between images.}
\label{fig:alpha}
\end{center}             
\end{figure}

To ensure the smoothness of transformation fields, a regularization term defined on the tangent space of diffeomorphisms (called velocity fields) is often introduced in registration models. Having such a regularity with proper model parameters is critical to registration performance because they greatly affect the estimated transformations. Either too large or small-valued regularity can not achieve satisfying registration results (as shown in Fig~.\ref{fig:alpha}). 
 Models of handling the regularity parameter mainly include (i) direct optimizing a Bayesian model or treating it as a latent variable to integrate out via Expectation Maximization (EM) algorithm~\citep{ allassonniere2007towards,allassonniere2008stochastic,zhang2013bayesian, wang2021bayesian}, (ii) exhaustive search in the parameter space~\citep{jaillet2005adaptive,valsecchi2013evolutionary,ruppert2017medical}, and (iii) utilizing parameter continuation methods~\citep{haber2000optimization, haber2006multilevel,mang2015inexact,mang2019claire}. Direct optimization approaches define a posterior of transformation fields that includes an image matching term as a likelihood and a regularization as a prior to support the smoothness of transformations~\citep{zollei2007marginalized,allassonniere2008stochastic,toews2009bayesian}. Estimating regularization parameters of these models using direct optimization is not straightforward due to the complex structure of the posterior distribution. Simpson et al. infer the level of regularization in small deformation registration model by mean-field VB inference~\citep{jordan1999introduction}, which allows tractable approximation of full Bayesian inference in a hierarchical probabilistic model~\citep{simpson2012probabilistic, simpson2015probabilistic}. However, these aforementioned algorithms are heavily dependent on initializations, and are prone to getting stuck in the local minima of high-dimensional and non-linear functions in the transformation space. A stochastic approximative expectation maximization (SAEM) algorithm~\citep{allassonniere2008stochastic} was developed to marginalize over the posterior distribution of unknown parameters using a Markov Chain Monte Carlo (MCMC) sampling method. Later, Zhang et al. estimate the model parameters of regularization via a Monte Carlo Expectation Maximization (MCEM) algorithm for unbiased atlas building problem~\citep{zhang2013bayesian}. A recent model of Hierarchical Bayesian registration~\citep{wang2021bayesian} further characterizes the regularization parameters as latent variables generated from Gamma distribution, and integrates them out by an MCEM method.  

Despite the achievement of the aforementioned methods, estimating the regularization parameter in a high-dimensional and nonlinear space of 3D MRIs (i.e., dimension is typically $128^3$ or higher) inevitably leads to expensive computational cost through iterative optimizations. To address this issue, we present a deep learning approach to fast predict registration parameters. While there exist learning-based registration models for transformations~\citep{krebs2019learning,balakrishnan2018unsupervised,biffi2020explainable}, we are particularly interested in learning the relationship between pairwise images and optimal regularizations of transformations via regression. In order to produce "ground truth" regularization parameters, we first introduce a low-dimensional Bayesian model of image registration to estimate the best regularity from the data itself. Following a recent work of~\citep{wang2019registration}, we construct a posterior distribution of diffeomorphic transformations entirely in a bandlimited space with much lower dimensions. This greatly reduces the computational cost of data generation in training. The theoretical tools developed in this work are generic to various deformable registration models, e.g, stationary velocity fields that remain constant over time~\citep{arsigny2006log}. The model recommends optimal registration parameters for registration in real-time and has great potential in clinical applications (i.e., image-guided navigation system for brain shift compensation during surgery~\citep{luo2018feature}). To summarize, our main contributions are three folds:
\begin{itemize}
\item To the best of our knowledge, we are the first to present a predictive regularization estimation method for diffeomorphic image registration through deep learning.
\item We develop a low-dimensional Bayesian framework in a bandlimited Fourier space to speed up the training data generation.
\item Our model significantly speeds up the parameter estimation, while maintaining comparable registration results.
\end{itemize}

The paper is organized as follows. In sec.~\ref{sec:background}, we lay out the basics of image registration optimization in the LDDMM framework. In sec.~\ref{sec:bayesian}, we first develop a low-dimensional posterior distribution that is parametrized by bandlimited velocity fields. We then estimate the regularization parameter by maximizing the posterior. In sec.~\ref{sec:netarch}, we design a deep convolutional neural network that takes an image pair as input and adaptively predicts the optimal smoothness level for registration. In sec.~\ref{sec:exp}, we validate our model on both 2D synthetic data and 3D brain MRI scans.

\section{Background: Fast LDDMM With Geodesic Shooting}
\label{sec:background}
In this section, we first briefly review a fast image registration algorithm FLASH in the setting of LDDMM with geodesic shooting~\citep{zhang2015fast,zhang2019fast}. The core idea of the FLASH is to reparameterize diffeomorphic transformations effectively in its tangent space (also known as velocity fields), where the signals are smooth without developing high frequencies in the Fourier space. This allows all computations of the original LDDMM with geodesic shooting~\citep{vialard2012,younes2009evolutions} to be carried out in the bandlimited space of velocity fields with much lower dimensions. As a result, the FLASH algorithm significantly speeds up diffeomorphic image registration with little to no loss of accuracy.

\subsection{Geodesic Shooting in Fourier Spaces}
Given time-dependent velocity field $\tilde{v}_t \in \tilde{V}$, the diffeomorphism $\phi^{-1}_t \in \widetilde{\Diff}(\Omega)$ in the finite-dimensional Fourier domain can be computed as
\begin{linenomath}
\begin{align}\label{eq:finalleftinvariantfft}
\frac{d \tilde{\phi}^{-1}_t }{dt}&=-\tilde{\mathcal{D}} \tilde{\phi}^{-1}_t \ast \tilde{v}_t,
\end{align}
\end{linenomath}
where $\tilde{\mathcal{D}} \tilde{\phi}^{-1}_t$ is a tensor product $\tilde{\mathcal{D}} \otimes \tilde{\phi}^{-1}_t$, representing the Fourier frequencies of a Jacobian matrix $\tilde{\mathcal{D}}$. The $\ast$ is a circular convolution with zero padding to avoid aliasing. We truncate the output of the convolution in each dimension to a suitable finite set to avoid the domain growing to infinity. 

Geodesic shooting algorithm states that the transformation can be uniquely determined by integrating a partial differential equation with a given initial velocity $v_0$ forward in time. In contrast to the original LDDMM that optimizes over a collection of time-dependent velocity fields, geodesic shooting estimates an optimal initial velocity $v_0$. In this work, we adopt an efficient variant of geodesic shooting defined in Fourier spaces. We first discretize the Jacobian and divergence operators using finite difference approximations (particularly the central difference scheme) and then compute their Fourier coefficients. Detailed derivations can be found in the appendix section of FLASH~\citep{zhang2019fast}. The Fourier representation of the geodesic constraint that satisfies Euler-Poincar\'e differential  (EPDiff) equation is~\citep{zhang2015fast}
\begin{linenomath}
\begin{align}\label{eq:epdiffleft}
    \frac{\partial \tilde{v}_t}{\partial t} =\ad^{\dagger}_{\tilde{v}_t} \tilde{v}_t  =-\tilde{\K}\left[(\tilde{\mathcal{D}} \tilde{v}_t)^T \star \tilde{\mathcal{L}}(\alpha)\tilde{v}_t  + \tilde{\nabla} \cdot \tilde{\mathcal{L}}(\alpha)\tilde{v}_t \otimes \tilde{v}_t) \right],
\end{align}
\end{linenomath}
where $\tilde{\L}$ is a symmetric, positive-definite differential operator that is a function of parameter $\alpha$ (details are in Sec.~\ref{sec:bayesian}). Here $\tilde{\K}$ is the inverse operator of $\tilde{\L}$, and $\star$ is the truncated matrix-vector field auto-correlation~\footnote{The output signal maintains bandlimited after the auto-correlation operates on zero-padded input signal followed by truncating it back to the bandlimited space.}. The $\ad^{\dagger}$ is an adjoint operator to the negative Jacobi–Lie bracket of vector fields, $\ad_{\tilde{v}} \tilde{w} = -[\tilde{v}, \tilde{w}]
= \tilde{\mathcal{D}}\tilde{v} \ast \tilde{w} - \tilde{\mathcal{D}}\tilde{w} \ast \tilde{v}$. The operator $\tilde{\nabla} \cdot$ is the discrete divergence (computed by summing the Fourier coefficients of different directions over in $\tilde{\mathcal{D}}$) of a vector field. 

\subsection{FLASH: Fast Image Registration}
Consider a source image $S$ and a target image $T$ as square-integrable functions defined on a torus domain $\Omega = \mathbb{R}^d / \mathbb{Z}^d$ ($S(x), T(x) : \Omega \rightarrow \mathbb{R}$). The problem of diffeomorphic image registration is to find the geodesic (shortest path) of diffeomorphic transformations
$\phi_t \in \Diff(\Omega): \Omega \rightarrow \Omega, t \in [0, 1]$, such that the deformed image $S \circ \phi^{-1}_1$ at time point $t=1$ is similar to $T$. 

The objective function can be formulated as a dissimilarity term plus a regularization that enforces the smoothness of transformations
\begin{linenomath}
\begin{align}
\label{eq:lddmm}
E(\tilde{v}_0) &= \frac{\gamma}{2}\text{Dist} (S \circ  \phi_1^{-1}, T) + \frac{1}{2}(\tilde{\L}(\alpha) \tilde{v}_0, \tilde{v}_0)
, \, s.t., \text{Eq}.\eqref{eq:finalleftinvariantfft} \, \& \, \text{Eq}.\eqref{eq:epdiffleft}.
\end{align} 
\end{linenomath}
The $\text{Dist}(\cdot, \cdot)$ is a distance function that measures the dissimilarity between images. Commonly used distance metrics include sum-of-squared difference ($L_2$ norm) of image intensities~\citep{beg2005computing}, normalized cross-correlation (NCC)~\citep{avants2008symmetric}, and mutual information (MI)~\citep{wells1996multi}. The $\phi_1^{-1}$ denotes the inverse of deformation that warps the source image $S$ in the spatial domain when $t = 1$. Here $\gamma$ is a weight parameter balancing between the distance function and regularity term. The distance term stays in the full spatial domain, while the regularity term is computed in bandlimited space.

\section{Our Model: Deep Learning for Regularity Estimation in Image Registration}

In this section, we present a supervised learning model based on CNN to predict the regularity of image registration for a given image pair. Analogous to ~\citep{yang2017quicksilver}, we run optimization-based image registration to obtain training data. We introduce a low-dimensional Bayesian model of image registration to produce appropriate regularization parameters for training. 

\subsection{Low-dimensional Bayesian Model of Registration}
\label{sec:bayesian}
In contrast to previous approaches, our proposed model is parameterized in a bandlimited velocity space $\tilde{V}$, with parameter $\alpha$ enforcing the smoothness of transformations.

Assuming an independent and identically distributed (i.i.d.) Gaussian noise on image intensities, we obtain the likelihood
\begin{linenomath}
\begin{equation}
\label{eq:likelihood}
p(T \, | \, S\circ \phi_1^{-1}, \sigma^2) = \frac{1}{(\sqrt{2 \pi} \sigma ^2 )^{M}} \exp{ \left(-\frac{1}{2\sigma ^2}|| S \circ \phi_1^{-1} -T ||_2^2 \right)},
\end{equation} 
\end{linenomath}
where $ \sigma^2 $ is the noise variance and $M$ is the number of image voxels. The deformation $\phi_1^{-1}$ corresponds to $\tilde{\phi}_1^{-1}$ in Fourier space via the Fourier transform $\mathcal{F}(\phi_1^{-1})=\tilde{\mathcal \phi}_1^{-1}$, or its inverse $\phi_1^{-1}=\mathcal{F}^{-1}(\tilde{\mathcal \phi}_1^{-1})$. The likelihood is defined by the residual error between a target image and a deformed source at time point $t=1$. We assume the definition of this distribution is after the fact that the transformation field is observed through geodesic shooting; hence is not dependent on the regularization parameters.

Analogous to~\citep{wang2018efficient}, we define a prior on the initial velocity field $\tilde{v}_0$ as a complex multivariate Gaussian distribution, i.e.,
\begin{linenomath}
\begin{align}
\label{eq:prior}
p({\tilde{v}}_0 | \alpha ) = \frac{1}{(2 \pi)^{\frac{M}{2}} | {\tilde{\L}}^{-1}(\alpha) |^{\frac{1}{2}}} \exp{ \left(-\frac{1}{2}({\tilde{\L}(\alpha)} {\tilde{v}}_0, {\tilde{v}}_0) \right)},
\end{align}
\end{linenomath}
where $|\cdot|$ is matrix determinant. The Fourier coefficients of $\tilde{\mathcal{L}}$ is, i.e.,
$\tilde{\L} = \left( \alpha \tilde{\mathcal{A}} + 1\right)^3$, $\tilde{\A}(\xi_1 , \ldots, \xi_d) = -2 \sum_{j = 1}^d \left(\cos (2\pi \xi_j) - 1 \right)$. Here $\tilde{\A}$ denotes a negative discrete Fourier Laplacian operator with a $d$-dimensional frequency vector $(\xi_1 , \ldots, \xi_d)$, where $d$ is the dimension of the bandlimited Fourier space.

Combining the likelihood in Eq.~\eqref{eq:likelihood} and prior in Eq.~\eqref{eq:prior} together, we obtain the negative log posterior distribution on the deformation
parameter parameterized by $\tilde{v}_0$ as
\begin{linenomath}
\begin{align}
\label{eq:poster}
-\ln \, p(\tilde{v}_0 \, | \, S, T, \sigma^2, \alpha) = & \frac{1}{2}(\tilde{\L} \tilde{v}_0, \tilde{v}_0) +
\frac{\| S \circ \phi_1^{-1} - T \|_2^2}{2\sigma^2}  - \frac{1}{2}\ln|\tilde{\L}| +2M \ln \sigma + M\ln(2\pi).
\end{align}
\end{linenomath}

Next, we optimize Eq.~\eqref{eq:poster} over the regularization parameter $\alpha$ and the registration parameter $\tilde{v}_0$ by maximum a posterior (MAP) estimation using gradient descent algorithm. Other optimization schemes, such as BFGS~\citep{polzin2016memory}, or the Gauss-Newton method~\citep{ashburner2011diffeomorphic} can also be applied. 

\textbf{Gradient of \pmb {$\alpha$}.} To simplify the notation, first we define $f(\tilde{v}_0) \triangleq -\ln \, p(\tilde{v}_0 \, | \, S, T,  \sigma^2, \alpha)$. Since the discrete Laplacian operator $\tilde{\L}$ is a diagonal matrix in Fourier space, its determinant can be computed as $\displaystyle \prod_{j=1}^d (\alpha \tilde{\A}_j + 1 )^3$. Therefore, the log determinant of $\tilde{\L}$ operator is 
\begin{linenomath}
\begin{equation}
\ln |\tilde{\L}| = 3 \sum_{j=1}^d (\alpha \tilde{\A}_j + 1 ). \notag
\end{equation}
\end{linenomath}

We then derive the gradient term $\nabla_{\alpha}f(\tilde{v}_0)$ as
\begin{linenomath}
\begin{align}
    \nabla_{\alpha}f(\tilde{v}_0) = -\frac{3}{2}[ \sum_{j=1}^d \frac{\tilde{\A}_j}{\alpha \tilde{\A}_j + 1} - \langle (\alpha \tilde{\A} +1)^{5}\tilde{\A} \tilde{v}_0, \tilde{v}_0 \rangle].
    \label{eq: alphagrad}
\end{align}
\end{linenomath}

\textbf{Gradient of \pmb {$\tilde{v}_0$}.}
We compute the gradient with respect to $\tilde{v}_0$ by using a forward-backward sweep approach developed in~\citep{zhang2017frequency}. Steps for obtaining the gradient $\nabla_{\tilde{v}_0}f(\tilde{v}_0)$ are as follows:
\begin{enumerate}[label=(\roman*)]
\item Forward integrating the geodesic shooting equation Eq.\eqref{eq:epdiffleft} to compute $\tilde{v}_1$,
\item Compute the gradient of the energy function Eq.~\eqref{eq:lddmm} with respect to $\tilde{v}_1$ at $t = 1$, 
\begin{linenomath}
\begin{align}
    \nabla_{\tilde{v}_1}f(\tilde{v}_0) = \tilde{\mathcal{L}}^{-1} (\alpha) \left( \frac{1}{\sigma^2}(S \circ \phi_1^{-1} - T) \cdot \nabla (S \circ \phi_1^{-1}) \right).
    \label{eq:gradvq1}
\end{align}
\end{linenomath}
\item Bring the gradient $\nabla_{\tilde{v}_1}f(\tilde{v}_0)$ back to $t=0$ by integrating adjoint Jacobi fields backward in time~\citep{zhang2017frequency},
\begin{linenomath}
\begin{equation} 
    \frac{d{\hat{v}}}{dt} = -\ad ^{\dagger}_{\tilde{v}}{\hat{h}}, \quad
    \frac{d{\hat{h}}}{dt} = -\hat{v} -\ad_{\tilde{v}} \hat{h} + \ad ^{\dagger}_{\hat{h}} \tilde{v},
    \label{eq:adjacobi}
    \end{equation}
\end{linenomath}
    where $\hat{v} \in V$ are introduced adjoint variables with an initial condition $\hat{h} = 0, \hat{v} =  \nabla_{\tilde{v}_1}f(\tilde{v}_0)$ at $t=1$. 
\end{enumerate}

A summary of the optimization is in Alg.~\ref{alg1}. It is worthy to mention that our low-dimensional Bayesian framework developed in bandlimited space dramatically speeds up the computational time of generating training regularity parameters by approximately ten times comparing with high-dimensional frameworks.
\begin{algorithm}[!htb]
\SetArgSty{textnormal}
\SetKwInOut{Input}{Input}
\SetKwInOut{Output}{Output}
\DontPrintSemicolon
\Input{source image $S$ and target image $T$, step size $\epsilon$, step size $\tau$, iterations $r$,  algorithm stop criterion rate $u$, counter $q$, minimal convergence iteration $q_{min}$  }
\Output{optimal smoothness level $\alpha^{opt}$, and registration solution $\tilde{v}_0$}
 \tcc{Low-dimensional MAP Estimation for $\alpha$ }
 \For{$i=1$ to $r$} 
 {
1. Compute gradient $\nabla_{\alpha}f(\tilde{v}_0)$ by Eq.~\eqref{eq: alphagrad};\\
\tcc{Update $\alpha$ when its gradient is greater than zero;}
2. \uIf{$\lVert\nabla_{\alpha}f(\tilde{v}_0) \rVert ^2 > 1e-6$ } 
{$ \alpha^{opt} \leftarrow\alpha^{opt}- \tau \nabla_{\alpha}f(\tilde{v}_0);$ }
\Else{break;}
3. Forward integrating the geodesic evolution equation in Eq.~\eqref{eq:epdiffleft} with initial velocity $\tilde{v}_0$;\\
4. Compute gradient $\nabla_{\tilde{v}_1}f(\tilde{v}_0) $ by Eq.~\eqref{eq:gradvq1} then backward integrating adjoint equations Eq.~\eqref{eq:adjacobi} to generate $\nabla_{\tilde{v}_0}f(\tilde{v}_0) $ at $t=0$; \\
\tcc{Update $\tilde{v}_0$ when its gradient is greater than zero;}
5. \uIf{$\lVert \nabla_{\tilde{v}_0}f(\tilde{v}_0) \rVert ^2 > 1e-6$ }
{$ \tilde{v}_0 \leftarrow\tilde{v}_0- \epsilon \nabla_{\tilde{v}_0 }f(\tilde{v}_0);$ }
\Else{break;}
 \tcc{Compute algorithm stop rate }
6. Compute total energy Eq.~\eqref{eq:poster} for $i-$th iteration as $\rm{Obj}_i$;\\
7. \uIf{$\frac{\rm{Obj}_i - \rm{Obj}_{i-1}}{\rm{Obj}_i} < u$ }
{ $ q = q + 1 $ ;}
\Else{continue;}
 \tcc{Convergence check}
8. \uIf{$ q \geq q_{min}$ }
{ break ;}
\Else{continue;}
} 
\textbf{End}
\caption{MAP of low-dimensional Bayesian registration model for training data generation.} \label{alg1}
\end{algorithm}


\subsection{Network Architecture}
\label{sec:netarch}
\begin{figure}[h]
\begin{center}
 \includegraphics[width=1.0\textwidth] {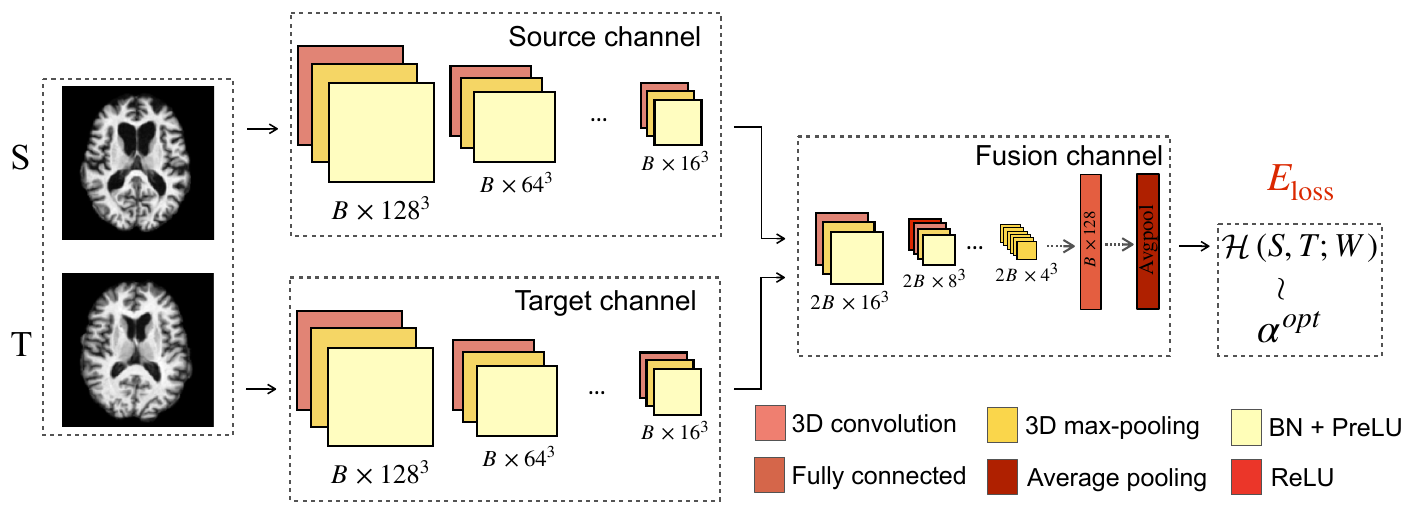}
     \caption{Illustration of our proposed network. From left to right: training data includes pairwise images, a CNN-based neural network, and a loss computed between the network output and the optimal parameter $\alpha^{opt}$, which is estimated from Alg.~\ref{alg1}.}
\label{fig:pipe}
\end{center}             
\end{figure}

Now we are ready to introduce our network architecture by using the estimated $\alpha$ and given image pairs as input training data. 
Fig.~\ref{fig:pipe} shows an overview flowchart of our proposed learning model. With the optimal registration regularization parameter $\alpha^{opt}$ obtained from an image pair (as described in Sec.~\ref{sec:bayesian}), a two-stream CNN-based regression network takes source images, target images as training data to produce a predictive regularization parameter. We optimize the network $\mathcal{H} (S, T; W)$ with the followed objective function,
\begin{linenomath}
\begin{align*}
   E_\text{loss} = \text{Err} [\mathcal{H} (S, T; W), \alpha^{opt}] + \text{Reg}(W),
\end{align*}
\end{linenomath}
where $\text{Reg}(W)$ denotes the regularization on convolution kernel weights $W$. $\text{Err}[\cdot,\cdot]$ denotes the data fitting term between the ground truth and the network output. In our model, we use $L_2$ norm for both terms. Other network architectures, e.g. 3D Residual Networks (3D-ResNet)~\citep{hara2017learning} and Very Deep Convolutional Networks (3D-VGGNet)~\citep{simonyan2014very,yang2018visual} can be easily applied as well. 

In our network, we input the 3D source and target images into separate channels that include four convolutional blocks. Each 3D convolutional block is composed of a 3D convolutional layer, a batch normalization ({BN}) layer with activation functions (PReLU or ReLU), and a 3D max-pooling layer. Specifically, we apply $5\times 5 \times 5$ convolutional kernel and $2\times 2 \times 2$ max-pooling layer to encode a batch (size as $B$) of source and target images ($128^3$) to  feature maps ($16^3 $). After extracting the deep features from source and target channels, we combine them into a fusion channel, which includes three convolutional blocks, a fully connected layer, and an average pooling layer to produce the network output.

\section{Experiment}
\label{sec:exp}
To demonstrate the effectiveness of the proposed low-dimensional Bayesian registration model, we validate it through three sets of experiments. For 2D synthetic data registration,  we deform a binary source image with velocity fields sampled from the prior distribution in Eq.~\eqref{eq:prior} with known regularization parameters to simulate target images. We show three convergence graphs of our MAP estimation and compare them with the ground truth parameters. 

Similarly, we synthesize $900$ image pairs using regularization parameters at different scales respectively, i.e., $\alpha = \{0.1, 1.0, 10.0\}$, to test the performance of our predictive model. We then use the predicted parameter $\alpha$ to run registration model and show the error maps between target and deformed images. 

For 3D brain MRI registration, we show results on both MAP and our network prediction. We first show the numerical difference between the MAP estimation and our prediction (i.e. predicted regularization parameter), and then report the mean error of deformed images between both methods across all datasets. We visualize the transformation grids and report the value of regularization parameters for both methods. To further investigate the accuracy of parameters generated by our model, we perform registration-based segmentation and examine the resulting segmentation accuracy over nine brain structures, including cortex, putamen, cerebellum, caudate, gyrus, brain stem, precuneus, cuneus, and hippocampus. We evaluate a volume-overlapping similarity measurement, also known as S$\o$rensen$-$Dice coefficient~\citep{dice1945measures}, between the propagated segmentation and the manual segmentation. The statistics of dice evaluation over $150$ registration pairs are reported. 

We last compare the computational efficiency on both time and memory consumption of the proposed method with a baseline model that performs Bayesian estimation of regularization parameter in the full spatial domain~\citep{zhang2013bayesian}.

To generate the training data of initial velocities, we run the proposed low-dimensional Bayesian registration algorithms until convergence. We use the Euler integrator in geodesic shooting and set the of integration steps as $10$. We set algorithm stop rate $u$ as $1e-6$ and minimal convergence iteration $q_{min}$ as 30. We use optimal truncated dimension for $\tilde{v}_0$ as 16 and $\sigma = 0.03$ according to~\citep{zhang2017frequency}. For the network setting, We initialize the convolution kernel weights using the He normal initializer~\citep{he2015delving} and use the Adam optimizer with a learning rate of $5e-4$ until convergence. We set $16$ and $1.0e-4$ as batch size and weight decay. The maximum epoch for 2D and 3D network training is $1000$.

\subsection{Data}
We run experiments on both 2D synthetic dataset and 3D real brain MRI scans. 

\noindent \textbf{2D synthetic data.} We generate synthetic bull-eye images with the size of $100 \times 100$ (as shown in Fig.~\ref{fig:2Dmapexp}). We manipulate the width $a$ and height $b$ of two ellipses by using equation $\frac{(x-50)^2}{a^2} + \frac{(y-50)^2}{b^2} = 1$.  

\noindent \textbf{3D brain MRIs.} We include $1500$ public T1-weighted brain MRI scans from Alzheimer's Disease Neuroimaging Initiative (ADNI) dataset~\citep{jack2008alzheimer}, Open Access Series of Imaging Studies (OASIS)~\citep{fotenos2005normative}, and LONI Probabilistic Brain Atlas Individual Subject Data (LPBA40)~\citep{shattuck2008construction}, among which $260$ subjects have manual delineated segmentation labels. All 3D data were carefully pre-processed as $128\times128\times128$, $1.25mm^{3}$ isotropic voxels, and underwent skull-stripped, intensity normalized, bias field corrected, and pre-aligned with affine transformation. 

For both 2D and 3D datasets, we split the images by using $70\%$ as training images, $15\%$ as validation images, and $15\%$ as testing images such that no subjects are shared across the training, validation, and testing stage. We evaluate the hyperparameters of models and generate preliminary experiments on the validation dataset. The testing set is only used for computing the final results.

\subsection{Results}
\begin{figure}[ht!]
\begin{center}
 \includegraphics[width=1.0\textwidth] {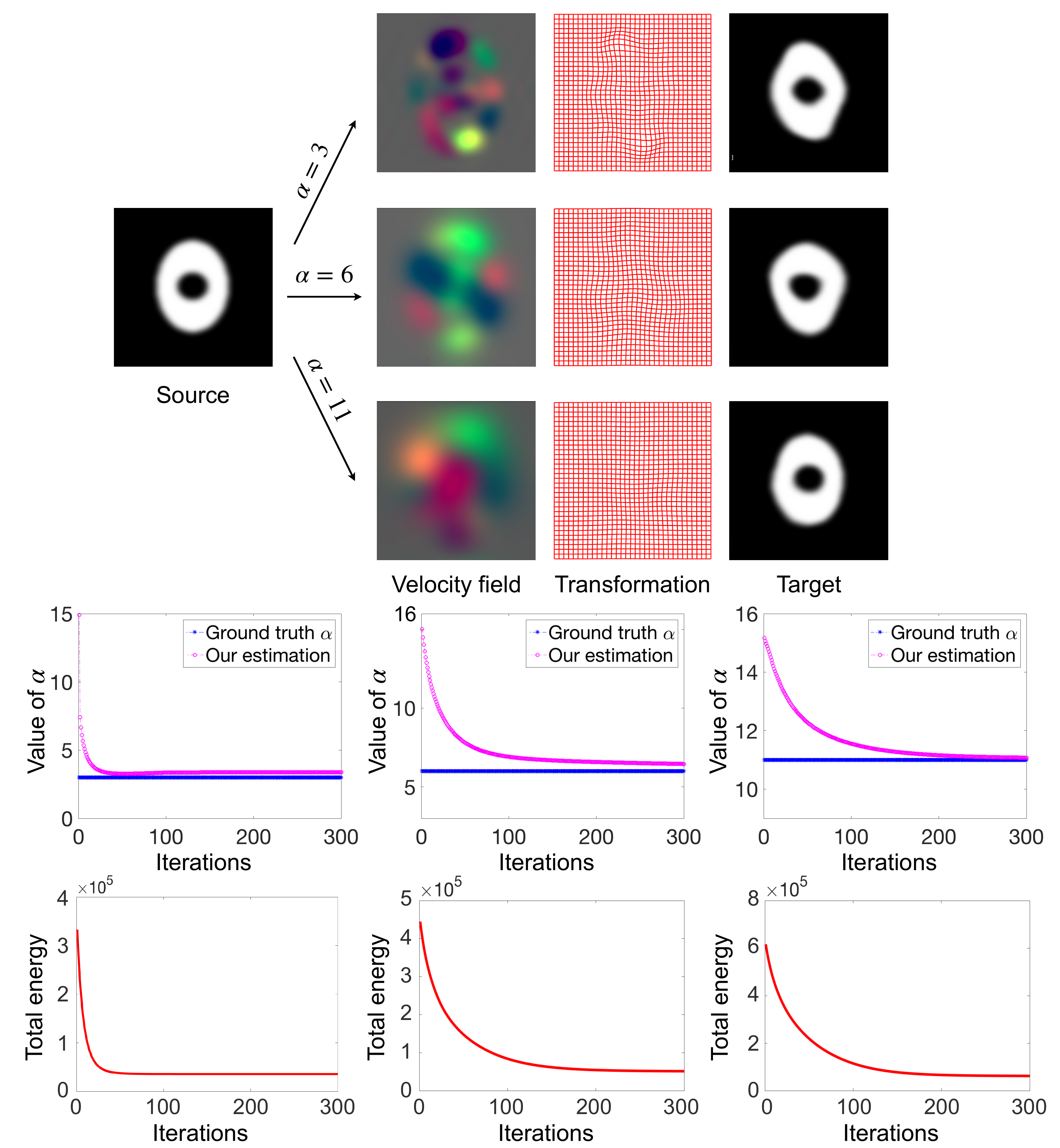}
     \caption{Top panel: source image, velocity fields generated from prior distribution and transformation fields (with known regularization parameter $\alpha = 3, 6, 11$), and target images produced by deformed source images; Middle panel: convergence graphs of estimated $\alpha$ by MAP for training data; Bottom panel: convergence graphs of total energy Eq.~\eqref{eq:poster}.}
\label{fig:2Dmapexp}
\end{center}             
\end{figure}

Fig.~\ref{fig:2Dmapexp} displays our MAP estimation of registration results including appropriate regularity parameters on 2D synthetic images. The middle panel of Fig.~\ref{fig:2Dmapexp} reports the convergence of $\alpha$ estimation vs. ground truth. It indicates that our low-dimensional Bayesian model provides trustworthy regularization parameters that are fairly close to ground truth for network training. The bottom panel of Fig.~\ref{fig:2Dmapexp} shows the convergence graph of the total energy for our MAP approach.

\begin{figure}[htb!]
\begin{center}
 \includegraphics[width=1.0\textwidth] {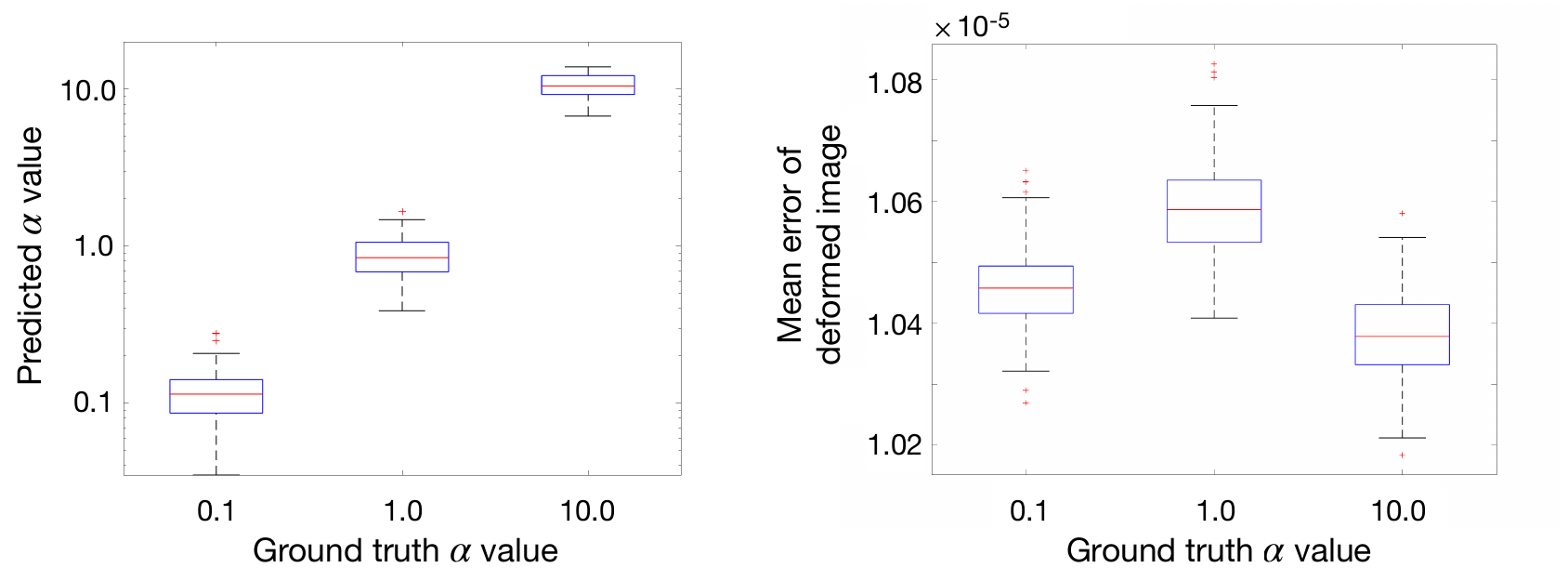}
     \caption{Left: our prediction of regularization parameters over $900$ image pairs synthesized with different ground truth parameters $\alpha = 0.1, 1.0, 10.0$; Right: error maps of image differences between deformed and target images.}
\label{fig:2Dnumerical}
\end{center}             
\end{figure}

Fig.~\ref{fig:2Dnumerical} further investigates the consistency of our network prediction. The left panel shows estimates of regularization parameter at multiple scales, i.e., $\alpha=0.1, 1.0, 10.0$, over $900$ 2D synthetic image pairs respectively. The right panel shows the mean error of image differences between deformed source images by transformations with predicted $\alpha$ and target images. While there are small variations on estimated regularization parameters, the registration results are very close (with averaged error at the level of $10^{-5}$). 

\begin{figure}[htb!]
\begin{center}
 \includegraphics[width=1.0\textwidth] {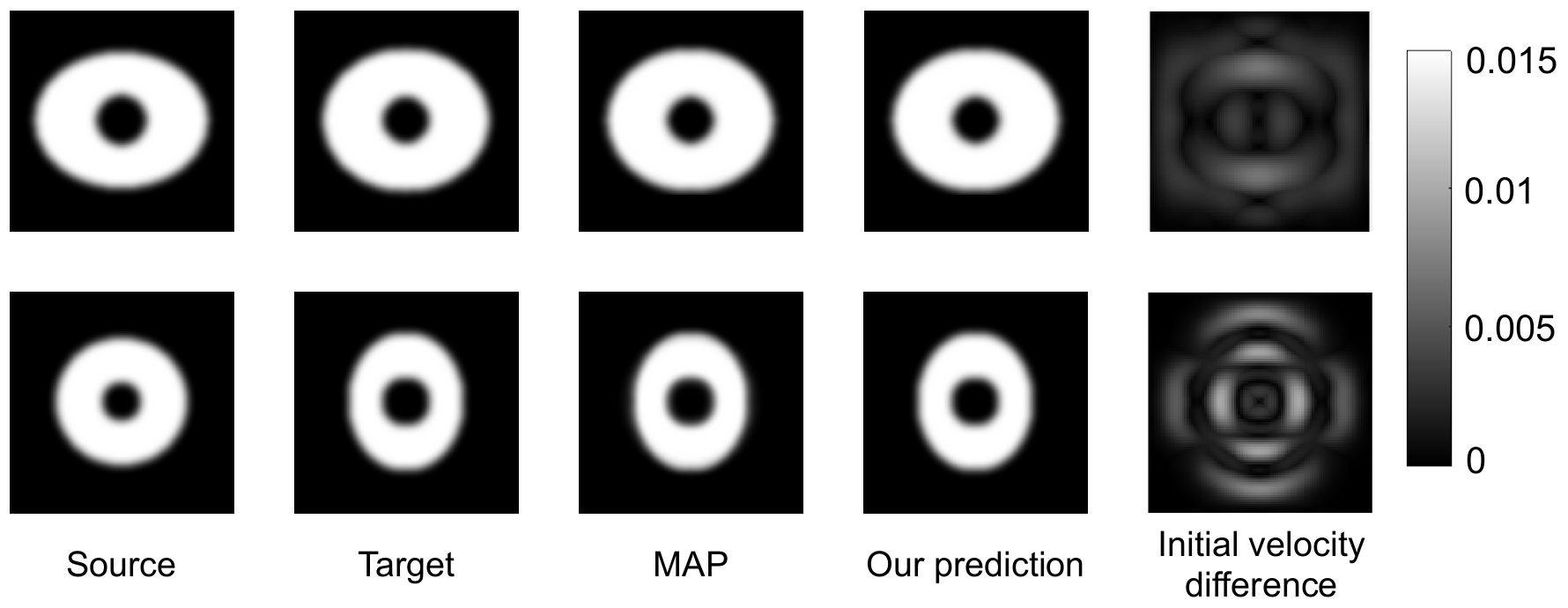}
     \caption{ Left to right: two examples of 2D synthetic source, target, deformed images by MAP and predictive deep learning method, and error maps of deformed image intensity differences.}
\label{fig:2Dresult}
\end{center}             
\end{figure}

Fig.~\ref{fig:2Dresult} shows examples of 2D pairwise image registration with regularization estimated by MAP and our predictive deep learning model. We obtain the regularization parameter $\alpha = 11.34$ (MAP) vs. $\alpha = 13.20$ (network prediction), and $\alpha = 5.44$ (MAP) vs. $\alpha = 6.70$ (network prediction). The error map of deformed images indicates that both estimations obtain fairly close registration results.   

\begin{figure}[!ht]
\begin{center}
 \includegraphics[width=.76\textwidth] {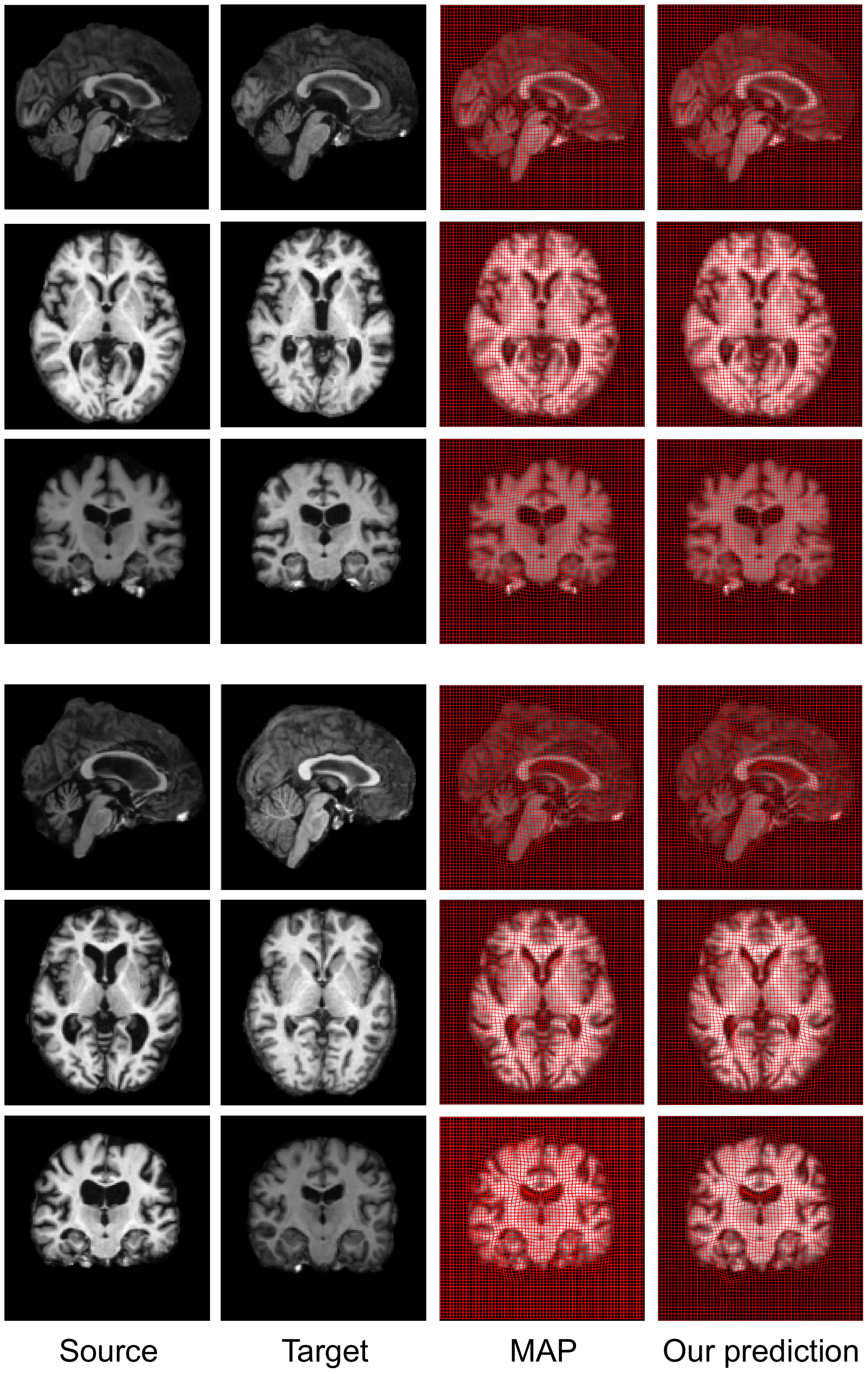}
     \caption{ Left to right: source, target, deformed images overlaid with the transformation grids generated by both methods. The optimal and predicted value of $\alpha$ for two showcases are $(8.90, 9.20)$ and $(3.44, 2.60)$.  }
\label{fig:3Dbrain}
\end{center}             
\end{figure}

Fig.~\ref{fig:3Dbrain} displays deformed 3D brain images overlaid with transformation grids for both methods. The parameter estimated from our model produces a comparable registration result. From our observation, a critical pattern between the optimal $\alpha^{opt}$ and its associate image pairs is that the value of $\alpha$ is relatively smaller when large deformation occurs. This is because the image matching term (encoded in the likelihood) requires a higher weight to deform the source image.

\begin{figure}[htb!]
\begin{center}
 \includegraphics[width=1.0\textwidth] {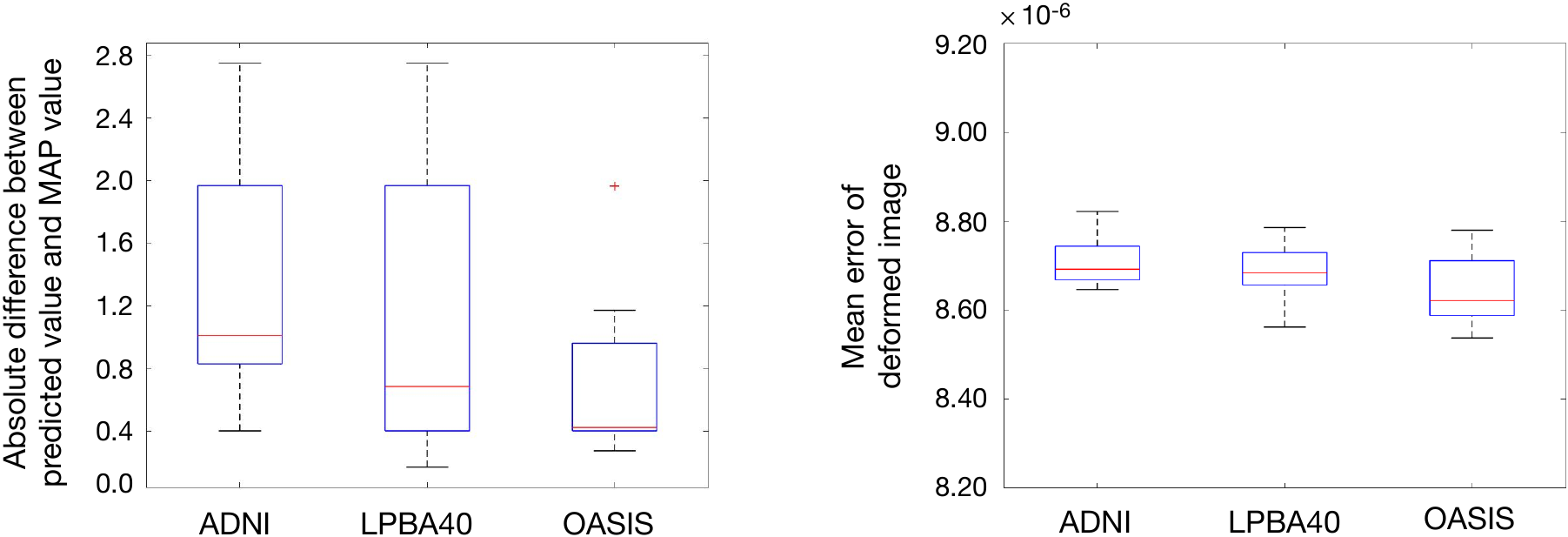}
     \caption{Left: statistics of the difference between our prediction and MAP estimation over $300$ image pairs from three data sets, ADNI, LPBA40, and OASIS; Right: error maps of image differences between deformed image by our prediction and MAP estimation.}
\label{fig:3Dnumerical}
\end{center}             
\end{figure}

Fig.~\ref{fig:3Dnumerical} investigates the consistency of our network prediction over three different datasets of 3D brain MRI. The left panel shows the absolute value of numerical differences between predicted regularization parameters and MAP estimations. The right panel shows the voxel-wise mean error of image differences between deformed images by transformations with predicted $\alpha$ and deformed images by MAP. While slight numerical difference on estimated regularization parameters exists, the 3D deformed images are fairly close (with averaged voxel-wise error at the level of $10^{-6}$).

\begin{figure}[!ht]
\begin{center}
 \includegraphics[width=.75\textwidth] {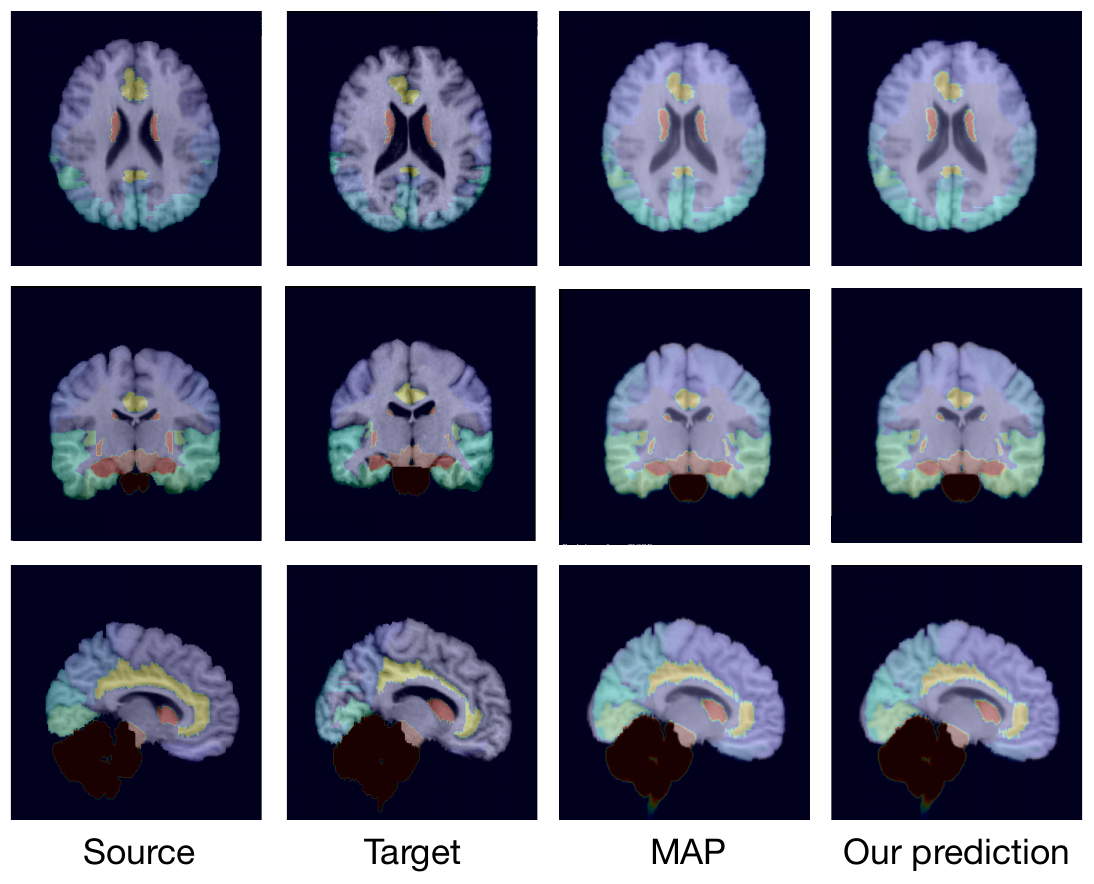}
     \caption{Axial, coronal and sagittal view of 3D segmentation labels with nine anatomical structures overlaid with source, target, deformed images by our low-dimensional MAP ($\alpha = 8.91$) and our network prediction ($\alpha = 6.80$).}
\label{fig:3Dseg}
\end{center}             
\end{figure}

Fig.~\ref{fig:3Dseg} visualizes three views of the deformed brain images (overlay with propagated segmentation label) that are registered by MAP and our prediction. Our method produces a registration solution, which is highly close to the one estimated by MAP. The propagated segmentation label fairly aligns with the target delineation for each anatomical structure. While we show different views of 2D slices of brains, all computations are carried out fully in 3D. 

\begin{figure}[!ht]
\begin{center}
 \includegraphics[width=1.0\textwidth] {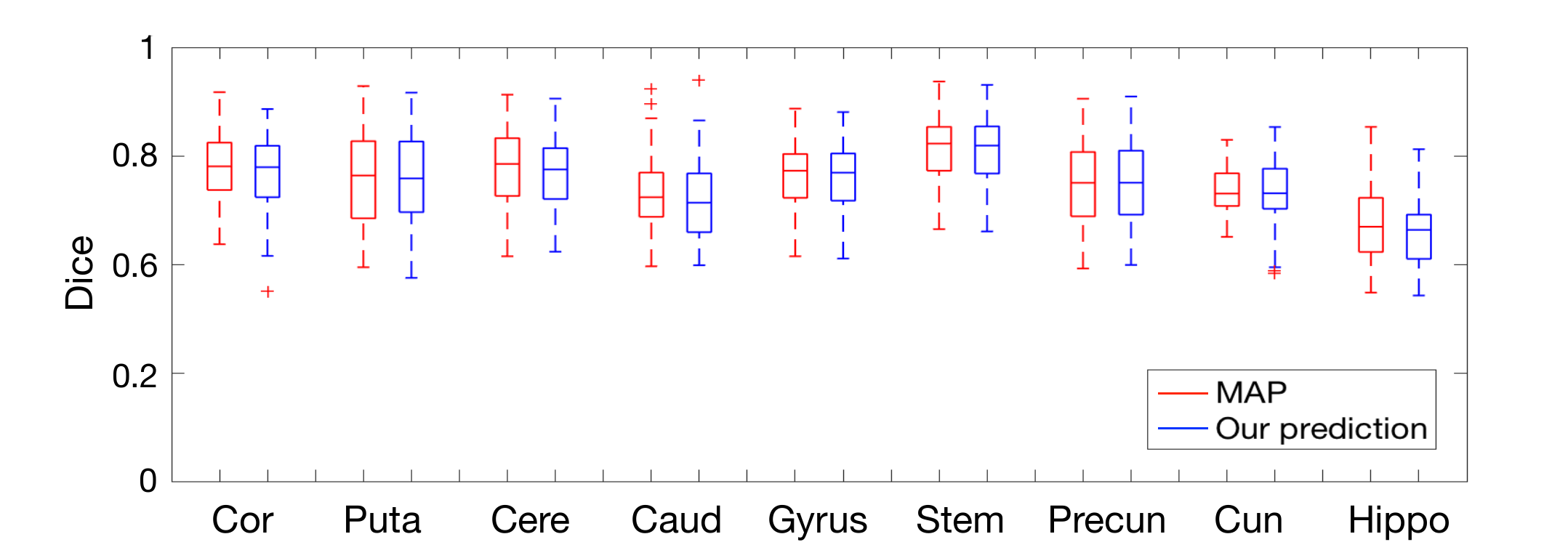}
     \caption{Dice scores of propagated manual segmentations for both methods for 150 registration pairs. Evaluations are performed over nine anatomical brain structures, Cor (cortex), Puta (putamen), Cere (cerebellum), Caud (caudate), gyrus, Stem (brain stem), Precun (precuneus), Cun (cuneus) and Hippo (hippocampus).}
\label{fig:3Ddice}
\end{center}             
\end{figure}

Fig.~\ref{fig:3Ddice} reports the volume overlapping of nine anatomical structures for both methods, including Cor(cortex), Puta (putamen), Cere (cerebellum), Caud (caudate), gyrus, Stem (brain stem), Precun (precuneus), Cun (cuneus), and Hippo (hippocampus). Our method produces comparable dice scores comparing with MAP estimations. This indicates that the segmentation-based registration by using our estimation achieves comparable registration performance with little to no loss of accuracy. 

Table.~\ref{Tab:time} quantitatively reports the averaged time and memory consumption of MAP estimation in full spatial image domain and our method. The proposed predictive model provides appropriate regularization parameters approximately $1000$ times faster than the conventional optimization-based registration method with a much lower memory footprint.
\begin{table}[htb!]
\caption{Time and memory consumption of MAP estimation of regularization in full-dimensional image space vs. our proposed low-dimensional Bayesian model, as well as network prediction.}
\centering
\begin{tabular}{|c|c|c|c|}
\hline
Methods       & Full-spatial  MAP & Low-dimensional MAP & Network Prediction \\ \hline
Runtime (Sec) & 1901             & 257                 & \textbf{2.16}               \\ \hline
Memory (MB)   & 450              & 119                 & \textbf{34.4}               \\ \hline
\end{tabular}
\label{Tab:time}
\end{table}

\section{Conclusion}
In this paper, we proposed a deep learning-based approach to model the relationship between the regularization of image registration and the input image data. We first developed a low-dimensional Bayesian model that defines image registration entirely in a bandlimited space. We then learned the mapping between regularization parameters and spatial images through a CNN-based neural network. To the best of our knowledge, we are the first to predict the optimal regularization parameter of diffeomorphic image registration by deep learning approaches. In contrast to existing methods, our developed model substantially improves the efficiency and the robustness of image registration. Our work has great potential in a variety of clinical applications, e.g. image-guided navigation system for neurosurgery in real-time. Potential future work may include: i) extending the current model to further consider adversarial examples, i.e., image outliers with significant differences; and ii) developing an unsupervised learning of registration parameter estimation to eliminate the need of training data generation (ground truth labels).

\acks{This work was supported by a startup funding at the University of Virginia.}

\ethics{The work follows appropriate ethical standards in conducting research and writing the manuscript, following all applicable laws and regulations regarding treatment of animals or human subjects.}

\newpage

\bibliography{paper}

\end{document}